\pdfoutput=1
\documentclass[11pt]{article}

\newif\ifaudioconsent
\audioconsenttrue

\newif\iffinalversion
\finalversiontrue

\usepackage{kezek}

\iffinalversion
  
\else
  
\fi

\newcommand{\kezek}{\textsc{Kezek}}
\newcommand{\db}{\,dB}

\title{Kezek: Splitting Mono Two-Person Conversations into Per-Speaker Tracks
for Full-Duplex Dialogue Data}
\author{Bagzhan Karl, Alizhan Karl}

\kztitle{\kezek{}: Splitting Mono Two-Person Conversations\\
into Per-Speaker Tracks for Full-Duplex Dialogue Data}
\kzauthors{%
  \begin{tabular}[t]{@{}l@{}}
    Bagzhan Karl\\[-1pt]
    {\small\color{kzgrey}\href{mailto:bagzhankarl@manifestro.io}{bagzhankarl@manifestro.io}}
  \end{tabular}\hspace{3.5em}%
  \begin{tabular}[t]{@{}l@{}}
    Alizhan Karl\\[-1pt]
    {\small\color{kzgrey}\href{mailto:alizhankarl@manifestro.io}{alizhankarl@manifestro.io}}
  \end{tabular}}
\kzaffil{Manifestro}
\kzlinks{%
  \href{https://github.com/Manifestro/kezek}{Code}%
  \quad\textperiodcentered\quad
  \href{https://huggingface.co/manifestro/kezek-v1}{Weights}%
  \quad\textperiodcentered\quad\href{https://manifestro.io/models/kezek}{Demo}}

\begin{document}
\kzmaketitle

\begin{abstract}
Full-duplex spoken dialogue models learn turn-taking, overlaps and
backchannels from conversations recorded with one channel per speaker, but
most available conversational audio is a single mono mix. We present
\kezek{}, a model that turns a mono recording of exactly two people into one
track per person with the original length and timing; overlaps and short
replies (``yeah'', ``mm-hm'') stay on the track of the person who said them.
\kezek{} decides \emph{who speaks when} and \emph{who is who} once for the
whole recording (stage~1: a two-voice activity model guided by a speaker
fingerprint trained to be insensitive to phase), and a conditioned
time-domain separator only carries out that decision (stage~2), with outputs
that always sum to the input. Standard separation metrics miss the failures
that matter for dialogue data, so we report short-reply retention, leakage
onto the other person's solo, channel swaps and coverage. Against the open pipeline
pyannote~$\rightarrow$~DialogueSidon on the same recordings, \kezek{} keeps
more short replies (0.83 vs.\ 0.69 on Open Yap test; 0.80 vs.\ 0.71 over the
other's speech on unseen telephone speech from CallHome English), mostly because
it returns the whole recording while the external pipeline drops part of it
(on the audio both return, reply retention is about the same: 0.859 vs.\ 0.865 on
Open Yap); it also leaks less, swaps channels less often, and runs 4--5 times
faster (about 230 hours of audio per GPU-hour). We also report negative
results and an ideal-mask ceiling that locate the remaining gap: quiet
replies under a louder voice. Code and weights are released under
permissive licences.
\end{abstract}

\section{Introduction}
\label{sec:intro}

Full-duplex spoken dialogue models listen and speak at the same time
\citep{nguyen2023dgslm,defossez2024moshi}. To learn when to interrupt, when
to back off and when to say ``mm-hm'' without taking the turn, they are
trained on conversations where each participant has a separate channel.
Such recordings are scarce. Most conversational audio (podcasts,
interviews, call recordings) exists only as a single mono mix.

Turning a mono mix into per-speaker tracks is usually done with a cascade:
a diarization system decides who speaks when, and a separation or
restoration model produces the tracks \citep{duplexchat2026}. For dialogue
data, three failure modes of such cascades matter more than average
separation quality:
\begin{enumerate}
  \item A channel swap puts a stretch of one person's speech on the other
        person's track, and the dialogue model learns the wrong turn-taking.
  \item Short replies get lost. Backchannels are short, quiet and often
        overlap the other speaker; they are exactly what a full-duplex model
        needs to learn, and exactly what diarization tends to drop.
  \item Leakage, the other person's voice audible on a track, teaches the
        model to treat its own channel as containing both speakers.
\end{enumerate}
A data-preparation tool must also return the \emph{whole} recording with
the original timing, and it must be fast enough to process tens of thousands
of hours.

\begin{kzhighlights}
{\sffamily\bfseries\color{kzaccent}Contributions}
\begin{itemize}
  \item A two-stage design in which \emph{who is who} is decided once for the
        whole recording and a conditioned separator only executes that
        decision, with an exact mixture-consistency constraint
        (Section~\ref{sec:method}).
  \item An evaluation protocol for dialogue-data separation: short-reply
        retention over speech and in pauses, leakage, per-recording channel
        swaps and coverage, applicable both with reference tracks and with
        utterance-level annotation only (Section~\ref{sec:metrics}).
  \item Two practical findings: energy-based activity labels of close-talk
        recordings mark the other speaker's crosstalk as speech, and fixing
        this with word-gated labels removes most channel swaps; and a raw-audio
        speaker fingerprint becomes competitive on conversational speech only
        once it is trained to ignore phase (Section~\ref{sec:data}).
  \item A comparison with the open pipeline pyannote~$\rightarrow$~DialogueSidon
        on Open Yap test and on unseen telephone speech, plus negative
        results and an ideal-mask ceiling that locate the remaining gap
        (Sections~\ref{sec:experiments}--\ref{sec:analysis}).
\end{itemize}
\end{kzhighlights}

\section{Related work}
\label{sec:related}

\paragraph{Speaker diarization.}
End-to-end neural diarization with a powerset output
\citep{plaquet2023powerset} underlies pyannote, the most widely used open
diarization toolkit. Target-speaker voice activity detection (TS-VAD)
\citep{medennikov2020tsvad} conditions activity estimation on speaker
profiles and handles overlap well. Both decide activity in local windows and
link windows by clustering speaker embeddings such as ECAPA-TDNN
\citep{desplanques2020ecapa}. A linking error is a channel swap.

\paragraph{Speech separation and target-speaker extraction.}
Time-domain masking networks such as Conv-TasNet \citep{luo2019convtasnet}
and time--frequency models such as TF-GridNet \citep{wang2023tfgridnet}
separate overlapping speech with high fidelity on synthetic mixtures.
Target-speaker extraction \citep{zmolikova2019speakerbeam} conditions the
separator on an enrollment of the target voice, often through FiLM
\citep{perez2018film}. Our separator is a Conv-TasNet-style model
conditioned on activity hints and on profiles of both speakers.

\paragraph{Data for full-duplex dialogue models.}
dGSLM \citep{nguyen2023dgslm} and Moshi \citep{defossez2024moshi} are trained
on two-channel conversations. DuplexChat \citep{duplexchat2026} builds such
data from podcasts with a cascade of pyannote diarization and the generative
restoration model DialogueSidon \citep{dialoguesidon}; channel swaps and
short replies are not measured there. We use that cascade as our external
baseline.

\section{Measuring dialogue-data separation}
\label{sec:metrics}

\paragraph{Task.}
Input: a mono recording at 16\,kHz of a conversation between exactly two
people, of any length, with no enrollment sample. Output: two tracks $y_A$,
$y_B$ of the same length and timing, each containing one person, and a
manifest with per-window confidence. Overlaps must be split, and every short
reply must be on the track of the person who produced it.

\paragraph{Metrics.}
Track order is fixed once per recording (the better of the two
permutations over the whole recording), as a downstream user would get it.
\begin{itemize}
  \item Short replies are utterances shorter than 1\,s. A reply is kept if
        its energy on the speaker's own track is within $-6$\db{} of the
        reference. We report replies over the other person's speech and in
        pauses separately; those over speech are the hard case (about two
        thirds of all replies in Open Yap).
  \item Leakage is the energy on a track during the \emph{other} person's
        solo, relative to that person's energy (lower is better).
  \item The no-swap rate is the share of recordings with no window
        (10\,s) assigned to the wrong person.
  \item Coverage is the share of the input duration returned; anything a
        system drops counts as lost.
  \item Overlap SI-SDRi \citep{leroux2019sisdr} is measured on frames where
        both speak, for systems whose output is sample-aligned with the input;
        for generative outputs it is not comparable.
  \item Speed is hours of audio processed per GPU-hour.
\end{itemize}
All energy-based metrics fit one gain per output track on that person's solo
frames, the same rule for every system, so that generative systems that
change the level are not penalised.
The energy criterion for replies is lenient on its own: the unprocessed
mixture copied to both tracks keeps 0.99 of replies (\cref{app:metrics}),
because it keeps everything. Reply retention is therefore meaningful only
read together with leakage, and every table here reports both.

\paragraph{Present and audible replies.}
The analysis of quiet replies (\cref{sec:analysis}) uses a stricter
criterion. A reply counts as kept only if it is \emph{present}: the
projection of the track onto the reply's reference has a coefficient of at
least 0.5 ($-6$\db{}); and \emph{audible}: the projected reply energy is at
least the energy of the rest of the track over the reply interval
(0\db{}). Under this criterion the reference tracks score 1.00, silence
0, and the unprocessed mixture 0.40 on replies over the other's speech.

\paragraph{Without reference tracks.}
For corpora with only utterance-level annotation (CallHome), a reply over
speech counts as kept when its track stands at least 6\db{} above the level
that leaks onto the same track during the other's solo within $\pm5$\,s.
Two sanity systems bound the scale: \emph{gated} (the mixture switched by the
annotation) scores 1.00, and \emph{mix} (the mixture on both tracks) scores
0.17 on replies over speech and 0\db{} leakage.

\section{Method}
\label{sec:method}

\subsection{Overview}
\begin{figure}[t]
  \centering
  \begin{tikzpicture}[
      font=\sffamily\small,
      >={Stealth[length=2.2mm]},
      box/.style={draw=kzaccent, line width=0.7pt, fill=white, rounded corners=2pt,
                  align=center, minimum height=11mm, inner sep=4pt},
      io/.style={draw=kzbase, fill=kzpaper, rounded corners=2pt, align=center,
                 minimum height=9mm, inner sep=4pt},
      arr/.style={->, draw=kzink, line width=0.6pt},
      lbl/.style={font=\sffamily\scriptsize, text=kzgrey, align=center},
      stage/.style={draw=none, fill=kzaccentlight, rounded corners=4pt},
      stagetitle/.style={font=\sffamily\scriptsize\bfseries, text=kzaccent}
    ]
    \node[io, minimum width=20mm] (x) {mono mix $x$\\[-1pt]{\scriptsize whole recording}};

    \node[box, right=11mm of x, yshift=11mm, minimum width=25mm] (fp)
      {speaker fingerprint\\{\scriptsize 10\,M}};
    \node[box, right=11mm of x, yshift=-11mm, minimum width=25mm] (pn)
      {PairNet\\{\scriptsize two-voice activity, 11.8\,M}};
    \draw[arr] (fp) -- node[lbl, right=1pt] {frame cosines\\to the 2 voices} (pn);

    \node[box, right=50mm of $(fp)!0.5!(pn)$, minimum width=25mm, minimum height=26mm] (sep)
      {KezekTSE-L\\{\scriptsize conditioned separator}\\{\scriptsize 19.6\,M}};

    \node[io, right=11mm of sep, yshift=7mm, minimum width=13mm] (ya) {$y_A$};
    \node[io, right=11mm of sep, yshift=-7mm, minimum width=13mm] (yb) {$y_B$};

    \begin{scope}[on background layer]
      \node[stage, fit=(fp)(pn), inner xsep=6mm, inner ysep=4mm,
            yshift=2.5mm, minimum height=0pt] (s1) {};
      \node[stage, fit=(sep), inner xsep=4mm, inner ysep=4mm, yshift=2.5mm] (s2) {};
    \end{scope}
    \node[stagetitle, anchor=north west] at ($(s1.north west)+(1.2mm,-0.6mm)$)
      {STAGE 1 \,·\, who speaks when};
    \node[stagetitle, anchor=north west] at ($(s2.north west)+(1.2mm,-0.6mm)$)
      {STAGE 2 \,·\, separate};

    \draw[arr] (x.east) -- ++(4mm,0) |- (fp.west);
    \draw[arr] (x.east) -- ++(4mm,0) |- (pn.west);
    \draw[arr] (fp.east) -- node[lbl, text=kzink, above] {profiles A, B} (fp.east -| sep.west);
    \draw[arr] (pn.east) -- node[lbl, text=kzink, below] {activity $a_A(t),\,a_B(t)$} (pn.east -| sep.west);
    \draw[arr, draw=kzgrey, dashed] (x.south) -- ++(0,-14mm) -| node[lbl, below, pos=0.25] {mixture waveform} (sep.south);
    \draw[arr] (sep.east) -- ++(4mm,0) |- (ya.west);
    \draw[arr] (sep.east) -- ++(4mm,0) |- (yb.west);

    \node[lbl, below=1.5mm of yb, text=kzaccent] {$y_A + y_B = x$};
  \end{tikzpicture}
  \caption{The \kezek{} pipeline. Stage~1 decides activity and identity once
  for the whole recording; stage~2 separates under that decision in 8\,s
  chunks and its outputs always sum to the input.}
  \label{fig:overview}
\end{figure}
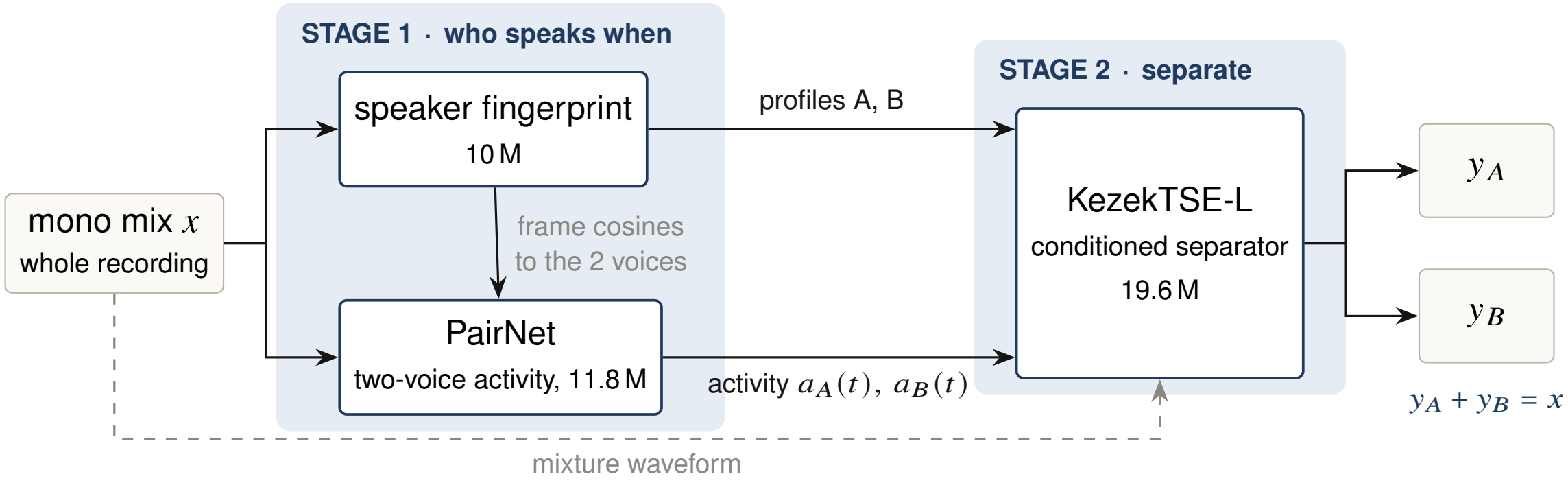

Earlier single-model designs that received a short window and had to decide
by themselves whether the target was present either suppressed a speaker on
real audio or collapsed into silence. \kezek{} therefore splits the problem:
stage~1 decides activity and identity for the whole recording; stage~2
separates, given that decision (Figure~\ref{fig:overview}).

\subsection{Speaker fingerprint}
\label{sec:fingerprint}
A 10.1\,M-parameter speaker embedding on raw 16\,kHz audio: a convolutional
front end (strides 5--4--4--4, one frame per 20\,ms, learned log
compression), a Conformer \citep{gulati2020conformer} (6 layers, 256-d),
attentive statistics pooling and a 192-d output. It replaces ECAPA-TDNN
\citep{desplanques2020ecapa} and runs 4.6$\times$ faster
(29\,539$\times$ vs.\ 6\,427$\times$ real time).

\paragraph{Training.} AAM-softmax (margin 0.2, scale 30) over 8\,961 people
from Libri-Light, VoxPopuli and Open Yap train, crops of 1--3\,s. Three
details target conversations. Each person is also present as ``siblings''
at speed factors 0.92 and 1.08, and these copies count as \emph{different}
classes, so the embedding must separate voices that differ only slightly. A
triplet term passes the anchor and a confusable partner (a sibling or the
nearest class, re-mined every 1\,000 steps) through the \emph{same}
recording chain and the positive through a different one, so that the
channel cannot be used as an identity cue. The recording chain includes a
telephone band of random shape and random all-pass cascades (1--4
second-order sections, 80--4000\,Hz) on 60\% of crops; without the latter,
the raw-waveform network was sensitive to phase alone
(\cref{sec:fp-analysis}).

\subsection{Stage 1: PairNet}
\label{sec:pairnet}
PairNet (11.8\,M parameters) estimates the activity of the two people over
the whole recording without window linking. Its encoder is the
fingerprint's raw-audio front end and Conformer, which produce, every
20\,ms, a 128-d voice embedding and a speech/overlap estimate. A log-mel
branch (80 bands, 20\,ms hop, per-recording normalisation, two
convolutions) is added to the encoder frames; its projection starts at
zero, so it can be added to a trained model without changing its output.

The two voices are found once for the whole recording with the fingerprint
(1.5\,s windows, 0.25\,s hop, two groups). Two Transformer decoder layers
shared by both voices compare each frame with the two voices and output
$a_A(t)$, $a_B(t)$; the decoder also receives, per frame, the cosine of the
fingerprint to each of the two voices. Inference runs on 10\,s windows with
2\,s of context. The identity of A and B is fixed by the recording-level
voices, so no linking is needed.

The loss is a permutation-invariant binary cross-entropy on activity, a
speech/overlap term and a voice term that pulls the found voices to the
labelled ones. Frames of utterances shorter than 1\,s are weighted
$\times$4. Training uses 40\,s crops: 70\% from Open Yap train with
word-gated labels (half of them centred on a real short reply) and 30\%
synthetic conversations from Libri-Light and Open Yap pieces with sibling
voices, short replies and overlaps, all through the recording chain.

Stage~1 processes about 2\,400 hours of audio per GPU-hour (the first
version, without the log-mel branch and fingerprint cosines, about
15\,900); the separator dominates the run time of the pipeline. Activity is thresholded at 0.3.

\subsection{Stage 2: KezekTSE-L separator}
A Conv-TasNet-style separator \citep{luo2019convtasnet} (19.6\,M parameters;
encoder $N{=}512$, kernel 32, stride 16, bottleneck 256, hidden 768, 8 blocks
$\times$ 3 repeats). Inputs: the mixture waveform; the activity hints
$a_A, a_B$ added at the bottleneck; and the profiles of both speakers via
FiLM \citep{perez2018film} in every block. It predicts two masks with a
softmax over sources and applies the projection
\begin{equation}
  y_i \leftarrow y_i + \tfrac{1}{2}\left(x - y_A - y_B\right),
  \qquad i \in \{A, B\},
\end{equation}
so that $y_A + y_B = x$ exactly. Nothing in the input is lost; it can only be
assigned to the wrong track.

\paragraph{Training targets.}
Reference for speaker $i$: that person's speech after the same recording
chain, plus background noise weighted by $a_i/(a_A + a_B)$ (0.5 when both are
silent), so that the references always sum to the mixture. The loss on each
track is
\begin{equation}
  \mathcal{L} = -\mathrm{SISDR}_{\text{act}}
    + 0.5\,\lvert\Delta L\rvert
    - 2\,\mathrm{SISDR}_{\text{ovl}}
    + \mathcal{L}_{\text{STFT}}
    + \mathcal{L}_{\text{reply}}
    + \mathcal{L}_{\text{leak}},
\end{equation}
with SI-SDR on the person's active frames (capped at 30\db{}), the level
error $\Delta L$ in dB, a separate SI-SDR on overlap frames, a
multi-resolution STFT L1 loss normalised by the mixture, the level error on
short replies, and a penalty on energy during the other person's solo
(floored at $-30$\db{}). The separate overlap term keeps overlaps from
being swamped by the far more frequent solo frames; on its own it was not
enough to make a separator learn overlaps from natural data, which led to the
curriculum in \cref{sec:data}. During training the activity hints are corrupted (edge
shifts of $\pm$40\,ms, 10\% of short segments dropped, 3\% false alarms,
occasionally all zero) so that the separator tolerates stage~1 errors.

\paragraph{Inference.}
The separator runs over the whole recording (8\,s chunks, 6\,s hop). A pass
restricted to the neighbourhood of overlaps kept fewer short replies at the
same leakage.

\section{How the models were trained}
\label{sec:data}

\paragraph{Corpora.}
Open Yap 1K \citep{openyap1k} is the conversational corpus: English
two-person conversations with a close-talk channel per speaker and
word-level transcripts. We split it by speaker, 1\,346 / 113 / 121
conversations (train / val / test), with no person in more than one split.
Libri-Light \citep{kahn2020librilight}, read speech of about 7\,400
speakers, is used for synthetic overlaps and for the fingerprint; VoxPopuli
\citep{wang2021voxpopuli} adds 1\,558 speakers (116\,h) for the
fingerprint. MUSAN \citep{snyder2015musan} and simulated room impulse
responses \citep{ko2017rir} supply the noise, music and reverberation of the
recording chain.

\paragraph{Recording chain.}
To match real recordings, training audio is passed, with probability 0.8,
through reverberation, noise and music, equalisation, compression, a
telephone band and a codec. Validation and test are evaluated both on the
clean sum of the two channels (\emph{clean}) and on a pre-computed render
through this chain (\emph{as recorded}).

\paragraph{Word-gated activity labels.}
Open Yap ships energy-based activity per channel. Close-talk microphones pick
up the other person: in frames where only the other person speaks, the
crosstalk on a headset channel has a median level of $-16$\db{} (10th--90th
percentile $-23$ to $-11$\db{}). This activity therefore marks crosstalk as
speech: 29\% of ``active'' frames contain none of the speaker's own words,
and about 80\% of labelled overlaps are not confirmed by the transcripts. A
first activity model trained on these labels swapped channels in 36--46\% of
conversations. Given the reference activity in place of its own prediction,
the same pipeline had no swaps at all (activity F1 0.993); the labels were
the problem, and speaker identification was working. We therefore
label a frame active only if it has energy \emph{and} lies within
$\pm$0.15\,s of one of the speaker's own words, and close pauses shorter than
0.3\,s inside speech; without the latter, 45\% of ``short replies'' are
fragments of ordinary turns. \Cref{tab:labels} shows the effect.

\begin{table}[t]
  \centering
  \caption{Effect of word-gated labels on the stage-1 activity model
  (prototype, WavLM features), Open Yap val, 113 conversations. Wrong track:
  share of replies assigned to the wrong person on real podcast dialogues,
  judged by speaker embedding against solo clips: 55 English replies from 10
  podcasts and 14 Kazakh replies from 5 podcasts.}
  \label{tab:labels}
  \small
  \renewcommand{\arraystretch}{1.15}
  \begin{tabular}{@{}lccccccc@{}}
    \toprule
    & \multicolumn{2}{c}{\sffamily Activity F1} & \multicolumn{2}{c}{\sffamily Short replies}
    & \multicolumn{2}{c}{\sffamily No-swap conv.} & \sffamily Wrong track \\
    \cmidrule(lr){2-3}\cmidrule(lr){4-5}\cmidrule(lr){6-7}\cmidrule(lr){8-8}
    \sffamily Labels & \sffamily clean & \sffamily rec & \sffamily clean & \sffamily rec
      & \sffamily clean & \sffamily rec & \sffamily EN / KK \\
    \midrule
    Energy VAD & 0.877 & 0.827 & 0.731 & 0.711 & 63.7\% & 54.0\% & 25.5\% / 28.6\% \\
    \kzrow Word-gated & \best{0.952} & \best{0.943} & \best{0.896} & \best{0.862}
      & \best{97.3\%} & \best{96.5\%} & \best{3.6\%} / \best{0\%} \\
    \midrule
    \textit{Reference activity} & \textit{0.993} & \textit{0.992} & --- & --- & \textit{100\%} & \textit{100\%} & --- \\
    \bottomrule
  \end{tabular}
\end{table}

The word-gated model also includes profile-conditioned heads (TS-VAD
style); on real dialogues it puts as few replies on the wrong track as
pyannote (3.6\% / 0\%).

\paragraph{Separator curriculum.}
Trained from scratch on natural Open Yap windows, separators learned to
route solo speech and did not start separating overlaps, which are short and
rare there. A TF-GridNet that could overfit 8 windows to +20.2\db{} stayed at
$-1.4$ to +0.4\db{} on validation overlaps. \Cref{tab:curriculum} shows the
path that worked: pre-train a medium model on long, clean synthetic overlaps
of Libri-Light speakers (4\,s windows, one speaker solo for 2--4\,s, the
other for 1--3\,s, overlap 1--3\,s at SIR $-5\ldots+5$\db{}, 20\% single
speaker); move to Open Yap with synthetic overlaps built from solos of
different people, natural windows and Libri-Light replay; then widen the
model to the large size by initialising it from the medium one. Training the
large model from scratch on the same mixture was far slower.

\begin{table}[t]
  \centering
  \caption{Separator training path. Window: SI-SDRi on Open Yap val overlap
  windows (clean / as recorded). Pipeline: overlap SI-SDRi of the whole
  pipeline on Open Yap val (clean / as recorded), each with the stage~1 of
  its time, so pipeline numbers compare only within a row's neighbours. The
  Libri-Light row's window numbers are zero-shot. All runs on one RTX PRO
  6000, batch 16, 4\,s crops. Highlighted: the separator used in \kezek{}.}
  \label{tab:curriculum}
  \small
  \renewcommand{\arraystretch}{1.15}
  \setlength{\tabcolsep}{4.5pt}
  \begin{tabular}{@{}>{\raggedright\arraybackslash}p{0.3\linewidth}crcccc@{}}
    \toprule
    \sffamily Stage & \sffamily Params & \sffamily Steps & \sffamily Synthetic
      & \sffamily Window & \sffamily Pipeline \\
    \midrule
    TF-GridNet S, Open Yap, scratch & 2.0\,M & 2.5\,k & --- & +0.37 / $-0.02$ & --- \\
    \midrule
    M, Libri-Light overlaps, scratch & 13.3\,M & 50\,k & +9.68 & +5.28 / +3.49 & --- \\
    M, Open Yap mixture & 13.3\,M & 17.5\,k & +9.98 & +7.55 / +5.09 & +9.49 / +7.01 \\
    L, Open Yap mixture, scratch & 19.6\,M & 12.5\,k & +4.74 & +2.73 / +1.86 & --- \\
    L, widened from M & 19.6\,M & 27.5\,k & +10.75 & +8.75 / +6.15 & +9.86 / +7.65 \\
    \kzrow L, on our fingerprint profiles & 19.6\,M & 15\,k & --- & +9.02 / +6.00 & +10.11 / +7.67 \\
    \bottomrule
  \end{tabular}
\end{table}

The medium model runs at 3.4 training iterations per second and the large at
2.1; at inference they process about 750 and 480 hours of audio per GPU-hour
on their own. The final step adapts the separator to the profiles of our own
fingerprint instead of ECAPA; it improves clean windows (+9.02 vs.\
+8.75\db{}, leak $-23.0$ vs.\ $-20.2$\db{}) at a small cost as recorded.

\paragraph{Licensing.}
\Cref{tab:licences} lists every corpus used and the terms it is used under.
Only data whose terms allow commercial use and distribution of trained
weights enters training; CallHome, which is non-commercial, is used for
evaluation only. No Open Yap audio or derived audio is redistributed.

\begin{table}[t]
  \centering
  \caption{Corpora and the terms they are used under.}
  \label{tab:licences}
  \small
  \renewcommand{\arraystretch}{1.2}
  \begin{tabular}{@{}l>{\raggedright\arraybackslash}p{0.29\linewidth}
                   >{\raggedright\arraybackslash}p{0.38\linewidth}@{}}
    \toprule
    \sffamily Corpus & \sffamily Used for & \sffamily Terms \\
    \midrule
    Open Yap 1K \citep{openyap1k} & training, validation, test
      & Data Use Agreement v1: commercial use and model distribution allowed;
        no redistribution of audio \\
    Libri-Light \citep{kahn2020librilight} & synthetic overlaps, fingerprint
      & LibriVox audiobooks (public domain); code MIT \\
    VoxPopuli \citep{wang2021voxpopuli} & fingerprint
      & data CC0; European Parliament material, source acknowledged \\
    MUSAN \citep{snyder2015musan} & noise and music in the recording chain
      & CC BY 4.0; content selected to allow commercial use \\
    RIRS\_NOISES \citep{ko2017rir} & reverberation
      & Apache 2.0 \\
    CallHome English \citep{canavan1997callhome,callhome_talkbank} & evaluation only
      & CC BY-NC-SA 4.0 (TalkBank) \\
    \bottomrule
  \end{tabular}
\end{table}

\section{Experiments}
\label{sec:experiments}

\subsection{Setup}
\paragraph{Systems.}
\kezek{} (version~1) is the fingerprint, PairNet stage~1 and KezekTSE-L,
41\,M parameters in total, activity threshold 0.3, separator over the whole
recording. The external system is pyannote~$\rightarrow$~DialogueSidon as
published in the DuplexChat code (30 sampling steps, fixed seed), run
without any change or tuning; its 24\,kHz output is resampled to 16\,kHz and
its global delay is compensated (residual $\leq$ 4\,ms). Whatever the
external pipeline drops is counted as lost and reported as coverage.

\paragraph{Evaluation sets.}
Open Yap test has 121 conversations, first 300\,s of each, clean and as
recorded, scored against the reference channels. CallHome English
\citep{canavan1997callhome} has 128 telephone calls with exactly two
annotated speakers (first 300\,s, 10.7\,h), unseen by both systems and
scored by the annotation (Section~\ref{sec:metrics}); it is used for
evaluation only. Speed is measured on one RTX PRO 6000 GPU.

\subsection{Main results}

\begin{table}[t]
  \centering
  \caption{\kezek{} vs.\ pyannote~$\rightarrow$~DialogueSidon on the same
  recordings, whole recordings. Open Yap test is \emph{as recorded}.
  Replies: share of short replies ($<$1\,s) kept on the right track; on Open
  Yap scored against the reference channels over all replies, on CallHome
  scored by the annotation, separately over the other's speech and in
  pauses. Leak: energy on a track during the other's solo.}
  \label{tab:main}
  \small
  \setlength{\tabcolsep}{5pt}
  \begin{tabular}{@{}llccccccc@{}}
    \toprule
    & & \multicolumn{3}{c}{\sffamily Replies kept $\uparrow$} & & & & \\
    \cmidrule(lr){3-5}
    \sffamily Set & \sffamily System & \sffamily all & \sffamily over speech & \sffamily in pause
      & \sffamily Leak (dB) $\downarrow$ & \sffamily No-swap $\uparrow$
      & \sffamily Coverage $\uparrow$ & \sffamily h\,/\,GPU-h $\uparrow$ \\
    \midrule
    Open Yap test & External & 0.687 & --- & --- & $-10.1$ & 0.793 & 0.735 & 51 \\
    \kzrow & \kezek{} & \best{0.830} & --- & --- & $\mathbf{-13.9}$ & \best{0.934} & \best{1.000} & \best{231} \\
    \midrule
    CallHome & External & --- & 0.711 & 0.798 & $-9.7$ & 0.797 & 0.816 & 46 \\
    \kzrow & \kezek{} & --- & \best{0.796} & \best{0.907} & $\mathbf{-11.8}$ & \best{0.859} & \best{1.000} & \best{232} \\
    \bottomrule
  \end{tabular}
\end{table}

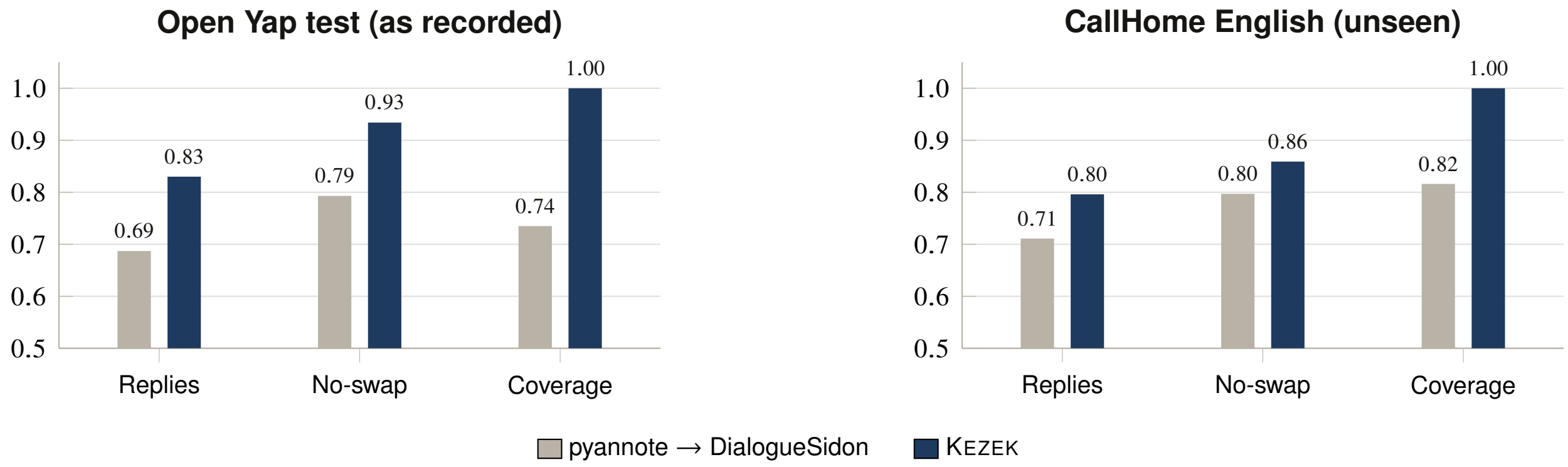
\begin{figure}[t]
  \centering
  \pgfplotsset{
    kzbars/.style={
      ybar=5pt, bar width=10pt, width=0.48\linewidth, height=46mm,
      scale only axis=false,
      ymin=0.5, ymax=1.05, ytick={0.5,0.6,0.7,0.8,0.9,1.0},
      symbolic x coords={Replies,No-swap,Coverage},
      xtick=data, enlarge x limits=0.25,
      tick label style={font=\sffamily\scriptsize},
      label style={font=\sffamily\scriptsize},
      title style={font=\sffamily\small\bfseries, text=kzink, yshift=-1mm},
      axis line style={draw=kzbase}, tick style={draw=kzbase},
      ymajorgrids, grid style={draw=kzbase!40},
      axis x line*=bottom, axis y line*=left,
      nodes near coords, every node near coord/.append style={
        font=\sffamily\tiny, text=kzink, /pgf/number format/fixed,
        /pgf/number format/fixed zerofill, /pgf/number format/precision=2},
      yticklabel style={/pgf/number format/fixed,
        /pgf/number format/fixed zerofill, /pgf/number format/precision=1},
      legend style={font=\sffamily\scriptsize, draw=none, fill=none,
        legend columns=2, /tikz/every even column/.append style={column sep=4mm}},
      legend image code/.code={\draw[#1] (0cm,-0.1cm) rectangle (0.25cm,0.1cm);},
    },
    kzext/.style={fill=kzbase, draw=none},
    kzours/.style={fill=kzaccent, draw=none},
  }
  \begin{tikzpicture}
    \begin{axis}[kzbars, title={Open Yap test (as recorded)}, legend to name=kzleg]
      \addplot[kzext]  coordinates {(Replies,0.687) (No-swap,0.793) (Coverage,0.735)};
      \addplot[kzours] coordinates {(Replies,0.830) (No-swap,0.934) (Coverage,1.000)};
      \legend{pyannote $\rightarrow$ DialogueSidon, \kezek{}}
    \end{axis}
  \end{tikzpicture}\hfill
  \begin{tikzpicture}
    \begin{axis}[kzbars, title={CallHome English (unseen)}]
      \addplot[kzext]  coordinates {(Replies,0.711) (No-swap,0.797) (Coverage,0.816)};
      \addplot[kzours] coordinates {(Replies,0.796) (No-swap,0.859) (Coverage,1.000)};
    \end{axis}
  \end{tikzpicture}\\[1mm]
  \ref*{kzleg}
  \caption{Main comparison at a glance. ``Replies'': all short replies on Open
  Yap, replies over the other's speech on CallHome. \kezek{} also runs 4--5$\times$ faster
  (231 vs.\ 51 hours of audio per GPU-hour on Open Yap).}
  \label{fig:main}
\end{figure}

Table~\ref{tab:main} and Figure~\ref{fig:main} show that \kezek{} keeps more short replies, leaks less,
swaps channels less often and returns the whole recording on both sets.
The result holds on CallHome, a telephone domain that neither system was
trained on. On the part of each recording that the external pipeline does
return, its reply retention is close to \kezek{}'s (Open Yap: 0.865 vs.\
0.859; CallHome: 0.814 over speech), but its leakage remains worse
($-8.5$ vs.\ $-13.3$\db{} on Open Yap). Most of the external pipeline's loss
therefore comes from what its diarization filter drops; the rest from
leakage and swaps. Both systems swap more on telephone speech (\kezek{}: 14\%
of calls) than on Open Yap (7\%).

\kezek{} is 4--5 times faster (last column of Table~\ref{tab:main}). At 230 hours per
GPU-hour, 100\,k hours take about 430 GPU-hours.

\subsection{Own components vs.\ a WavLM + ECAPA prototype}
\label{sec:prototype}
Before version~1 we built a prototype with the same separator but a stage~1
on frozen WavLM Base+ features \citep{chen2022wavlm}, window linking with
ECAPA embeddings, and an ensemble of four TS-VAD-style heads.
\Cref{tab:prototype} compares the two once on Open Yap test. Version~1 has
no frozen third-party components (41\,M parameters, all trained by us,
against about 140\,M for the prototype, 115\,M of them frozen), is twice as
fast, and is better on clean overlaps; as recorded, it leaks more and keeps
fewer replies. The stage~1 used in this comparison was the predecessor of
the one in \cref{tab:main}; the gap in replies is what the final stage~1
narrows (\cref{tab:s1-conditions}).

\begin{table}[t]
  \centering
  \caption{Version~1 (own components) vs.\ the prototype (frozen WavLM
  Base+ and ECAPA), Open Yap test, run once. Version~1 here uses the stage~1
  that preceded the final one. Speed and parameters: prototype / version~1.}
  \label{tab:prototype}
  \small
  \renewcommand{\arraystretch}{1.15}
  \begin{tabular}{@{}lcccc@{}}
    \toprule
    & \multicolumn{2}{c}{\sffamily Clean} & \multicolumn{2}{c}{\sffamily As recorded} \\
    \cmidrule(lr){2-3}\cmidrule(lr){4-5}
    \sffamily Metric & \sffamily Prototype & \sffamily Version 1 & \sffamily Prototype & \sffamily Version 1 \\
    \midrule
    Overlap SI-SDRi (dB) $\uparrow$ & 10.08 & \best{10.49} & \best{8.57} & 8.41 \\
    Solo SI-SDR (dB) $\uparrow$ & \best{21.59} & 21.46 & \best{19.41} & 19.30 \\
    Leak (dB) $\downarrow$ & \best{$-20.06$} & $-19.58$ & \best{$-16.96$} & $-15.63$ \\
    Short replies kept $\uparrow$ & \best{0.900} & 0.864 & \best{0.871} & 0.811 \\
    \midrule
    Speed (h / GPU-h) $\uparrow$ & \multicolumn{4}{c}{118 / \best{231}} \\
    Parameters (M) & \multicolumn{4}{c}{$\approx$\,27 + 115 frozen / 41} \\
    \bottomrule
  \end{tabular}
\end{table}

\subsection{Stage 1 under recording conditions}
\Cref{tab:s1-conditions} evaluates stage~1 alone on 40 Open Yap val
conversations (300\,s each) under each effect of the recording chain.
PairNet matches the prototype's activity F1 everywhere and keeps more
replies in noise; elsewhere it keeps 1.3--2 points fewer. Splitting the
replies by context locates the gap: in pauses PairNet is slightly better
(0.911 vs.\ 0.904), over the other person's speech it is worse (0.853 vs.\
0.876). The development of PairNet raised replies under the full chain from
0.779 (first version) to 0.857 (short replies weighted $\times$4) and 0.873
(log-mel branch and crops around real replies).

\begin{table}[t]
  \centering
  \caption{Stage~1 alone under each recording condition: activity F1 and
  share of short replies found on the right person (40 Open Yap val
  conversations). Prototype: WavLM + ECAPA + 4-head ensemble.}
  \label{tab:s1-conditions}
  \small
  \renewcommand{\arraystretch}{1.15}
  \begin{tabular}{@{}lcccccc@{}}
    \toprule
    & \sffamily clean & \sffamily full chain & \sffamily reverb & \sffamily noise & \sffamily codec & \sffamily telephone \\
    \midrule
    \multicolumn{7}{@{}l}{\sffamily\itshape\color{kzgrey} Activity F1} \\
    Prototype & 0.938 & \best{0.928} & \best{0.925} & \best{0.920} & \best{0.939} & \best{0.937} \\
    \kzrow PairNet & 0.938 & 0.924 & 0.923 & 0.919 & 0.937 & 0.932 \\
    \multicolumn{7}{@{}l}{\sffamily\itshape\color{kzgrey} Short replies} \\
    Prototype & \best{0.904} & \best{0.886} & \best{0.873} & 0.846 & \best{0.901} & \best{0.893} \\
    \kzrow PairNet & 0.886 & 0.873 & 0.853 & \best{0.860} & 0.884 & 0.873 \\
    \bottomrule
  \end{tabular}
\end{table}

\subsection{Stage 1 as a speaker diarizer}
\label{sec:der}
Stage~1 alone is a two-speaker diarizer, so we also score it with the
standard diarization error rate (DER, overlap scored, pyannote.metrics) next
to pyannote speaker-diarization-community-1 \citep{plaquet2023powerset}, the
diarization front end of the external pipeline (\cref{tab:der}). References
are the word-gated activity of each Open Yap channel and the annotated
utterances of CallHome.

\begin{table}[t]
  \centering
  \caption{Diarization error rate (\%, collar 0; collar 0.25\,s in
  brackets) split into missed speech, false alarm and speaker confusion, with
  the Jaccard error rate (JER) and the share of recordings where exactly two
  speakers are found. Stage~1 of \kezek{} vs.\ pyannote community-1.}
  \label{tab:der}
  \small
  \renewcommand{\arraystretch}{1.15}
  \setlength{\tabcolsep}{4.5pt}
  \begin{tabular}{@{}llccccccr@{}}
    \toprule
    \sffamily Set & \sffamily System & \sffamily DER $\downarrow$ & \sffamily Miss & \sffamily FA
      & \sffamily Conf. $\downarrow$ & \sffamily JER $\downarrow$ & \sffamily 2 spk. & \sffamily h\,/\,GPU-h \\
    \midrule
    Open Yap, clean & pyannote & 25.11 (19.65) & 17.78 & 4.29 & 3.03 & 27.69 & 0.835 & 150 \\
    \kzrow & \kezek{} stage 1 & \best{15.27} (10.87) & 3.04 & 11.71 & \best{0.51} & \best{15.84} & \best{1.000} & $\approx$2\,400 \\
    \midrule
    Open Yap, as rec. & pyannote & 25.79 (20.51) & 16.14 & 5.21 & 4.44 & 29.30 & 0.851 & 154 \\
    \kzrow & \kezek{} stage 1 & \best{19.31} (14.42) & 3.23 & 15.26 & \best{0.82} & \best{18.94} & \best{1.000} & $\approx$2\,600 \\
    \midrule
    CallHome & pyannote & \best{18.44} (13.05) & 10.48 & 4.33 & 3.64 & \best{23.00} & 1.000 & 153 \\
    \kzrow & \kezek{} stage 1 & 30.05 (20.42) & 5.06 & 23.00 & \best{1.99} & 27.94 & 1.000 & $\approx$2\,300 \\
    \bottomrule
  \end{tabular}
\end{table}

On Open Yap, stage~1 has a lower DER than pyannote (15.3 vs.\ 25.1 clean,
19.3 vs.\ 25.8 as recorded) and always finds exactly two speakers. On
CallHome it is worse (30.1 vs.\ 18.4). Its speaker confusion is low on every
set (0.5--2.0\% against 3.0--4.4\%); nearly all of its error is false-alarm
speech. That comes from the low activity threshold (0.3) and from closing
short pauses in the labels, both chosen to keep short replies. With a
threshold of 0.7, DER falls to 14.8\% on Open Yap as recorded and 23.1\% on
CallHome; this was measured for sensitivity only, and \kezek{} keeps 0.3.
On CallHome the false alarm stays near 10\% even at 0.7, and part of it is
probably the reference itself: utterance-level segments without annotated
backchannels and laughter. Stage~1 runs at about 2\,400 hours of audio per
GPU-hour.

\subsection{Overlap separation}
On the same Open Yap test run, \kezek{} reaches an overlap SI-SDRi of
+10.61\db{} on clean audio and +8.18\db{} as recorded; short replies 0.878
and 0.830; recordings without swaps 0.967 and 0.934; and leakage (energy of
the error on the other's solo) of $-19.5$ and $-14.9$\db{}. The external
pipeline's output is generative and not sample-aligned, so its SI-SDRi
($-29$\db{}) is not meaningful and is not compared.

\ifaudioconsent
\subsection{Listening examples}
Excerpts from Open Yap conversations of the public sample that \kezek{}
did not see in training (mono input, the two \kezek{} tracks and the
original channels) are available at \url{https://manifestro.io/models/kezek}. Open Yap audio is
reproduced with the written consent of The Agentic Data Company.
\fi

\section{Analysis}
\label{sec:analysis}

All numbers in this section are on Open Yap \emph{val}, as recorded, end to
end, unless stated otherwise.

\paragraph{Where replies are lost.}
\Cref{tab:attribution} swaps components between version~1 and the prototype
on 20 val conversations. With the prototype's stage~1 under our
fingerprint and separator, the pipeline keeps as many replies as the whole
prototype on clean audio (0.896) and more overlap and less leakage; with our
stage~1 it keeps 0.045 fewer. As recorded, the loss splits into $-0.026$ from
stage~1 and $-0.033$ from the separator and profiles. The replies that
stage~1 misses are absent from its output altogether, and lowering the
threshold does not bring them back: recovering short bursts down to an
activity of 0.15 adds 0.007, and a separator pass without activity hints
finds no more. Stage~1 is not the only source of loss, though: for quiet
replies the separator loses many even when given labelled activity (see
the ideal-mask ceiling below).

\begin{table}[t]
  \centering
  \caption{Swapping components (Open Yap val, 20 conversations, 300\,s
  each). Stage~1 of version~1 here is the predecessor of the final one.}
  \label{tab:attribution}
  \small
  \renewcommand{\arraystretch}{1.15}
  \begin{tabular}{@{}lcccccc@{}}
    \toprule
    & \multicolumn{3}{c}{\sffamily Clean} & \multicolumn{3}{c}{\sffamily As recorded} \\
    \cmidrule(lr){2-4}\cmidrule(lr){5-7}
    \sffamily Pipeline & \sffamily Overlap & \sffamily Leak & \sffamily Replies
      & \sffamily Overlap & \sffamily Leak & \sffamily Replies \\
    \midrule
    Prototype & +9.77 & $-21.9$ & \best{0.896} & +7.69 & \best{$-20.8$} & \best{0.851} \\
    Prototype stage 1 + our fingerprint and separator & \best{+10.26} & \best{$-22.0$} & \best{0.896} & \best{+7.88} & $-20.2$ & 0.818 \\
    \kzrow Version 1 & +10.11 & $-21.7$ & 0.851 & +7.67 & $-20.4$ & 0.792 \\
    Version 1 + short-burst rescue & +10.14 & $-21.6$ & 0.857 & +7.66 & $-20.1$ & 0.799 \\
    \bottomrule
  \end{tabular}
\end{table}

\paragraph{Ideal-mask ceiling.}
To see whether a masking separator can recover a reply from under a louder
voice at all, we apply an ideal ratio mask (STFT, 512-sample window, mixture
phase) and count replies over the other's speech by how much quieter they
are than the other voice (\cref{tab:irm}). This analysis uses the stricter
\emph{present and audible} reply criterion defined in \cref{sec:metrics},
under which the unprocessed mixture keeps only 0.396 of such replies. Replies louder than the other voice or up to
10\db{} quieter are nearly solved; more than 10\db{} below, the ideal mask
keeps 0.743 and \kezek{} 0.171. Even given labelled activity our separator
keeps only 0.286. On all 113 val conversations, where this bin holds 151
replies instead of 35, the same separator keeps 0.371 and the ideal mask
0.788; fine-tuning the separator on synthetic replies 5--25\db{} below the
main voice raised 0.371 by just 0.026. The remaining headroom is real but
bounded, and is not a matter of the level of replies in the training data.

\begin{table}[t]
  \centering
  \caption{Replies over the other's speech kept, by level relative to the
  other voice (20 Open Yap val conversations of 300\,s, as recorded; clean
  in brackets). A reply counts as
  kept only if it is present and audible (\cref{sec:metrics}); only replies
  at least 6\db{} above the crosstalk in their own reference are scored.}
  \label{tab:irm}
  \small
  \renewcommand{\arraystretch}{1.15}
  \begin{tabular}{@{}lcccc@{}}
    \toprule
    \sffamily System & \sffamily All & \sffamily Louder & \sffamily 0 to $-10$\db{} & \sffamily Below $-10$\db{} \\
    \midrule
    \textit{Ideal ratio mask} & \textit{0.938 (0.944)} & \textit{1.000 (1.000)} & \textit{1.000 (1.000)} & \textit{0.743 (0.784)} \\
    \kzrow \kezek{} version 1 & 0.722 (0.804) & 0.965 (0.968) & 0.827 (0.907) & 0.171 (0.405) \\
    Separator + labelled activity & 0.757 (0.804) & 0.947 (1.000) & 0.865 (0.860) & 0.286 (0.405) \\
    Mixture on both tracks & 0.396 (0.441) & 0.982 & 0.019 & 0.000 \\
    \bottomrule
  \end{tabular}
\end{table}

\begin{table}[t]
  \centering
  \caption{Things that did not help (Open Yap val, as recorded). Replies in the rows on activity hints and on synthetic quiet replies use the present-and-audible criterion (\cref{sec:metrics}); the other rows use the energy criterion.}
  \label{tab:negative}
  \small
  \renewcommand{\arraystretch}{1.25}
  \begin{tabular}{@{}>{\raggedright\arraybackslash}p{0.43\linewidth}>{\raggedright\arraybackslash}p{0.53\linewidth}@{}}
    \toprule
    \sffamily Change & \sffamily Result (vs.\ the configuration it was compared with) \\
    \midrule
    TF-GridNet \citep{wang2023tfgridnet} (2\,M) from scratch on natural windows
      & learns to route solo, overlap SI-SDRi $\leq$ +0.4\db{} \\
    Separator 19.6\,M $\rightarrow$ 26\,M
      & +7.68 vs.\ +7.67\db{} overlaps, leak unchanged, replies 0.805 vs.\ 0.792 \\
    Separator without activity hints
      & replies over speech 0.662 vs.\ 0.708, leak $-11.2$ vs.\ $-19.9$\db{},
        swaps in 4 of 20 recordings vs.\ 0 \\
    Separator fine-tuned on hints of the real stage~1, leak weight $\times$3.3
      & leak +0.35\db{} better, overlaps $-0.6$\db{}, replies $-0.04$ \\
    Synthetic replies 5--25\db{} below the main voice
      & quiet replies 0.397 vs.\ 0.371 (ideal mask 0.788; all 113 val conversations) \\
    Short-reply loss weight $\times$3 in the separator
      & replies 0.825 vs.\ 0.838, overlaps $-0.3$\db{} \\
    Self-supervised pre-training of the stage~1 encoder ($\sim$900\,h)
      & replies 0.75--0.81 vs.\ 0.82--0.85 \\
    \bottomrule
  \end{tabular}
\end{table}

\paragraph{Negative results.}
Table~\ref{tab:negative} lists changes that did not move the pipeline.
Capacity, loss weights and the level of replies in training data are not
the bottleneck. The activity hint cannot be dropped: without it the
separator starts suppressing real solo speech and swapping channels.

\subsection{Speaker fingerprint: phase sensitivity}
\label{sec:fp-analysis}
\Cref{tab:fp-series} follows the fingerprint from its first version to the
one used in \kezek{}. EER is measured on Open Yap val crops of 1.5--3\,s
through the full evaluation chain (reverberation, noise, music, equalisation,
compression, microphone bleed, telephone band, real MP3/Opus coding), on
Libri-Light val, and on hard pairs (same sex, top quarter of similarity).

\begin{table}[t]
  \centering
  \caption{Speaker fingerprint EER (lower is better) through its
  development. Each row starts from the weights of the row above.
  ``Hard'': mean of the hard-pair EER on Open Yap and Libri-Light.}
  \label{tab:fp-series}
  \small
  \renewcommand{\arraystretch}{1.15}
  \begin{tabular}{@{}l>{\raggedright\arraybackslash}p{0.36\linewidth}cccc@{}}
    \toprule
    & \sffamily Change & \sffamily Open Yap & \sffamily Libri & \sffamily Libri hard & \sffamily Hard \\
    \midrule
    ECAPA-TDNN & --- & \best{11.7} & 6.9 & 11.2 & 15.0 \\
    \midrule
    \#1 & raw audio, siblings, channel-matched pairs & 15.3 & 7.7 & 10.8 & 15.2 \\
    \#2 & + full chain, 25\% conversational anchors & 15.7 & 6.5 & 9.4 & 15.0 \\
    \#3 & + telephone band on 35\% of crops & 14.9 & 6.3 & --- & 14.7 \\
    \#4 & + telephone band of random shape & 14.6 & 5.9 & --- & 14.3 \\
    \#5 & + random all-pass on 60\% of crops & 13.6 & 5.2 & 7.2 & 13.2 \\
    \kzrow \kezek{} & + 1\,558 VoxPopuli voices & 13.9 & \best{5.0} & \best{6.8} & \best{13.2} \\
    \bottomrule
  \end{tabular}
\end{table}

\begin{table}[t]
  \centering
  \caption{Phase sensitivity of the raw-audio fingerprint, EER on Open Yap
  val where half of the crops pass through the named filter. The all-pass
  filter changes only the phase.}
  \label{tab:fp-phase}
  \small
  \renewcommand{\arraystretch}{1.15}
  \begin{tabular}{@{}lccc@{}}
    \toprule
    \sffamily Filter on half of the crops & \sffamily ECAPA
      & \sffamily Ours, \#4 & \sffamily Ours, \#5 (+ all-pass training) \\
    \midrule
    telephone band, zero phase  & 9.2 & \best{8.5} & --- \\
    telephone band, IIR (with phase) & \best{9.1} & 14.1 & 10.8 \\
    all-pass only ($|H|=1$) & \best{7.5} & 9.6 & 8.0 \\
    telephone band, resampled to 8\,kHz & \best{10.8} & 14.6 & 11.2 \\
    \bottomrule
  \end{tabular}
\end{table}

\paragraph{Finding.} The same telephone band costs our first raw-audio
fingerprint 8.5\% EER when applied with zero phase and 14.1\% with the phase
response of an IIR filter (\cref{tab:fp-phase}). A pure all-pass filter,
which leaves the magnitude spectrum unchanged, raises its EER to 9.6\%
(ECAPA: 7.5\%). Training with random all-pass cascades closes most of this
gap. A front end that is blind to phase by construction (a learned complex
filterbank followed by the modulus) removes the sensitivity (telephone
10.3\% vs.\ 10.8\% for ECAPA) but is coarser on clean conversational speech
(9.5\% vs.\ 7.5\% for the convolutional front end), so we kept the convolutional front end with all-pass
augmentation.

\paragraph{Where it stands.} The final fingerprint is better than ECAPA on
read speech and on hard pairs, and 4.6$\times$ faster, but still worse on
conversational Open Yap under the full chain (13.9\% vs.\ 11.7\%). Adding
1\,558 VoxPopuli speakers did not change that number, so the remaining gap
is not the number of training voices. What \kezek{} needs from the
fingerprint is to tell similar voices apart, and there it is ahead: on hard
Libri-Light pairs its EER is 6.8\% against 11.2\% for ECAPA.

\section{Limitations}
\label{sec:limitations}

Most of the short replies \kezek{} still loses are quiet replies under a
louder voice (Section~\ref{sec:analysis}).

\kezek{} handles exactly two speakers, offline, with no enrollment and no
speech restoration: unlike DialogueSidon, it does not clean the audio, it
only splits it.

Training and quantitative evaluation are in English. On a small set of five
Kazakh podcasts, the stage~1 activity model put none of 14 judged replies on
the wrong person (\cref{tab:labels}), but this sample is far too small for a
conclusion. Kazakh examples are shown on the demo page
(\url{https://manifestro.io/models/kezek}) for listening only; a
quantitative evaluation on Kazakh conversations is future work.

The Open Yap test split was evaluated more than once during development
(each time with the model choice fixed beforehand on val). CallHome was
evaluated once, and its scoring relies on utterance-level annotation with
no reference tracks.

\paragraph{Ethics.}
\kezek{} only reassigns audio that is already in a recording; it does not
synthesise speech. Separated tracks of a person's voice can still be misused
for voice cloning; users must have the right to process the recordings.
Following the Open Yap data use agreement, we do not try to identify
speakers and do not release Open Yap audio or derived tracks.

\section{Conclusion}
\label{sec:conclusion}

\kezek{} decides who is who once per recording and separates under that
decision, with outputs that sum to the input. Against the open
pyannote~$\rightarrow$~DialogueSidon cascade on the same recordings, in
domain and on telephone speech neither system saw in training, it returns
the whole recording, keeps more short replies over it (mostly because the
external pipeline drops part of the audio), leaks less, swaps channels less
often and runs 4--5 times faster. What remains is quiet replies under a
louder voice. Both stages lose them: some are missing from the stage~1
output, and for replies more than 10\db{} below the other voice even the
separator given labelled activity keeps 0.286, where an ideal mask keeps
0.743.

\paragraph{Availability.}
\begin{itemize}
  \item Code (MIT): \url{https://github.com/Manifestro/kezek}
  \item Weights (Apache-2.0): \url{https://huggingface.co/manifestro/kezek-v1}
\end{itemize}
\ifaudioconsent
Listening examples: \url{https://manifestro.io/models/kezek}.
\fi

\section*{Acknowledgements}
We thank The Agentic Data Company for Open Yap 1K and for permission to
publish the listening examples. VoxPopuli audio: we acknowledge the European
Parliament for creating and sharing these materials. The MUSAN corpus is used
under CC BY 4.0.

\bibliographystyle{plainnat}
{\small\bibliography{refs}}

\appendix
\section{Model details}
\label{app:details}

\Cref{tab:components} lists the three components of \kezek{} version~1
and \cref{tab:schedules} the training schedule of each.

\begin{table}[H]
  \centering
  \caption{Components of \kezek{} version~1. All are trained by us; nothing
  is frozen from third-party models.}
  \label{tab:components}
  \small
  \renewcommand{\arraystretch}{1.15}
  \begin{tabular}{@{}l>{\raggedright\arraybackslash}p{0.52\linewidth}r@{}}
    \toprule
    \sffamily Component & \sffamily Architecture & \sffamily Params \\
    \midrule
    Fingerprint & conv front end (strides 5-4-4-4) $\rightarrow$ Conformer
      (6 layers, 256-d) $\rightarrow$ attentive statistics pooling
      $\rightarrow$ 192-d & 10.1\,M \\
    PairNet & fingerprint-style encoder + log-mel branch $\rightarrow$
      128-d voice embedding per 20\,ms $\rightarrow$ 2 Transformer decoder
      layers & 11.8\,M \\
    KezekTSE-L & Conv-TasNet: $N{=}512$, $L{=}32$, stride 16, $B{=}256$,
      $H{=}768$, $P{=}3$, $X{\times}R = 8{\times}3$, gLN, FiLM by both
      profiles in every block, hint via 1$\times$1 conv at the bottleneck,
      two softmax masks & 19.6\,M \\
    \bottomrule
  \end{tabular}
\end{table}

\begin{table}[H]
  \centering
  \caption{Training schedules. All runs: AdamW, bf16, one NVIDIA RTX PRO
  6000. Each row starts from the weights of the previous row of the same
  component. ``---'': not recorded.}
  \label{tab:schedules}
  \small
  \renewcommand{\arraystretch}{1.15}
  \begin{tabular}{@{}>{\raggedright\arraybackslash}p{0.34\linewidth}cccccc@{}}
    \toprule
    \sffamily Stage & \sffamily LR & \sffamily Warmup & \sffamily Steps & \sffamily Batch & \sffamily Crop \\
    \midrule
    \multicolumn{6}{@{}l}{\sffamily\itshape\color{kzgrey} Fingerprint (AAM-softmax, $m{=}0.2$, $s{=}30$)} \\
    first version & 1e-3 & 2\,000 & 24.8\,k & 128 pairs & 1--3\,s \\
    full chain, conversational anchors & 5e-4 & 500 & 12\,k & 128 pairs & 1--3\,s \\
    telephone band 35\% & 3e-4 & 300 & 6\,k & 128 pairs & 1--3\,s \\
    random telephone band & 3e-4 & 300 & 6\,k & 128 pairs & 1--3\,s \\
    random all-pass & 3e-4 & 300 & 10\,k & 128 pairs & 1--3\,s \\
    + VoxPopuli (final) & 5e-4 & 1\,000 & 25\,k & 128 pairs & 1--3\,s \\
    \multicolumn{6}{@{}l}{\sffamily\itshape\color{kzgrey} PairNet} \\
    first version & 1e-3 (enc.\ 3e-4) & --- & 20\,k & 12 & 40\,s \\
    short replies $\times$4 & 3e-4 & 300 & 10\,k & 12 & 40\,s \\
    fingerprint cosines & 3e-4 & --- & 8\,k & 12 & 40\,s \\
    log-mel branch, reply crops (final) & 3e-4 & 300 & 12\,k & 12 & 40\,s \\
    \multicolumn{6}{@{}l}{\sffamily\itshape\color{kzgrey} KezekTSE-L separator} \\
    M, Libri-Light overlaps & 1e-3 cosine & 2\,000 & 50\,k & 16 & 4\,s \\
    M, Open Yap mixture & 5e-4 cosine & 500 & 17.5\,k & 16 & 4\,s \\
    L, widened from M & 5e-4 cosine & 500 & 27.5\,k & 16 & 4\,s \\
    L, fingerprint profiles (final) & 3e-4 & 500 & 15\,k & 16 & 4\,s \\
    \bottomrule
  \end{tabular}

  \smallskip
  {\footnotesize\raggedright Training speed: fingerprint $\approx$7.7 steps/s
  at every stage (first version 0.94\,h); PairNet 7.7 steps/s for the first
  version (0.72\,h) and 2.9 steps/s for the final one with the log-mel branch
  and fingerprint cosines (1.15\,h).\par}
\end{table}

\section{Metric definitions}
\label{app:metrics}

\paragraph{With reference tracks (Open Yap).}
\begin{itemize}
  \item Track order is chosen once per recording, as the better of
        the two permutations by envelope energy against the reference
        activity. The same rule applies to every system.
  \item One gain per output track is fitted by energy on that
        person's solo frames, so a system that changes the overall level is
        not penalised.
  \item Short replies are runs of a person's activity shorter than
        1\,s. A reply is kept if the energy of the output track over it is
        within $-6$\db{} of the reference.
  \item Leak is the energy of the output track during the other
        person's solo, relative to that person's energy. Because close-talk
        references themselves contain crosstalk, the reference tracks used
        as output score $-15.3$\db{} (clean) and $-15.7$\db{} (as
        recorded); this is the floor of the metric.
  \item Swaps are counted in 10\,s windows where both people have
        solo speech; a window is swapped if its best order differs from the
        recording's order. Windows a system returns as silence are skipped.
  \item The global delay of a system's output is found on
        the envelope and compensated; a residual above 20\,ms would stop the
        evaluation. The external pipeline's residual was at most 4\,ms.
\end{itemize}

\paragraph{Present and audible replies.}
Used for the quiet-reply analysis. Let $\alpha$ be the coefficient of the
projection of the output track onto the reply's reference over the reply
interval. The reply is present if $\alpha \geq 0.5$ and audible if the
energy of the projected part is at least the energy of the remainder of the
track over the same interval. Replies are split by the level of the reply
relative to the other voice: louder, 0 to $-10$\db{}, below $-10$\db{}.
Only replies at least 6\db{} above the crosstalk in their own reference are
scored (144 of 154 over speech, as recorded). Known answers on 20 val
conversations: reference tracks 1.000, silence 0, unprocessed mixture
0.441 / 0.396 on the scored replies over speech (clean / as recorded; 0.435
/ 0.403 counting all of them).

\paragraph{Without reference tracks (CallHome).}
A reply over speech counts as kept when its track stands at least 6\db{}
above that track's own level during the other person's solo within
$\pm$5\,s; a reply in a pause, when it is within $-6$\db{} of the mixture.
Leak is the track's energy on the other's solo relative to the mixture.

\begin{table}[H]
  \centering
  \caption{Known-answer checks run before every comparison. A metric is
  accepted only if these systems score as expected. The mixture on both
  tracks keeps nearly all replies under the energy criterion but has no
  separation (leak near 0\db{}), which is why replies are always reported
  with leakage.}
  \label{tab:gates}
  \small
  \renewcommand{\arraystretch}{1.15}
  \begin{tabular}{@{}llcccc@{}}
    \toprule
    \sffamily Set & \sffamily System & \sffamily Replies & \sffamily Leak (dB) & \sffamily No-swap & \sffamily Delay \\
    \midrule
    Open Yap test & reference tracks as output & 1.000 & $-15.3$ / $-15.7$ (floor) & 1.000 & 0\,ms \\
                  & mixture on both tracks & 0.994 / 1.000 & $-0.9$ / $-0.3$ & --- & --- \\
    CallHome & mixture gated by the annotation & 1.000 / 1.000 & $-262$ & 1.000 & --- \\
             & mixture on both tracks & 0.167 / 1.000 & 0.0 & --- & --- \\
    \bottomrule
  \end{tabular}

  \smallskip
  {\footnotesize Open Yap: clean / as recorded. CallHome replies: over
  speech / in a pause.}
\end{table}

Before the external comparison, the earlier evaluation of version~1 was
also reproduced exactly with the new scoring code (replies 0.864 /
0.811, leak $-19.58$ / $-15.63$\db{}, clean / as recorded), so that numbers
from different experiments are comparable.

\ifaudioconsent
\section{Listening examples}
\label{app:audio}
The examples come from the public Open Yap 1K sample and from
conversations that are in none of our training, validation or test splits.
We confirmed against the training data of every component that none of
these conversations was used in training. The four excerpts (30\,s each,
pseudonymous conversation IDs, excerpt number after the underscore):
\texttt{5a798c657bb0\_1}, \texttt{31bedfc99b53\_1},
\texttt{e11edcca8521\_2}, \texttt{2e4aec4666d8\_1}.
Each comes with the mono mixture, the two \kezek{} tracks and the original
channels.
Reproduced with the written consent of The Agentic Data Company.
\fi

\section{Complete experiment log}
\label{app:log}

Every experiment of the development of \kezek{} version~1, grouped by
component and in the order they were run within each group, with the conclusion recorded when it was closed. Each experiment
fixed its hypothesis, stopping gates and an outcome$\rightarrow$action table
before it started. Unless stated otherwise, numbers are on Open Yap val;
``clean'' and ``rec'' are the clean and as-recorded conditions. Omitted:
data downloads, listening-only checks, checks on private recordings, and one
experiment on a different task (suppressing a speaker from an enrollment
sample).

{\small
\renewcommand{\arraystretch}{1.18}
\setlength{\tabcolsep}{4pt}
\begin{longtable}{@{}>{\sffamily\bfseries\color{kzaccent}}l
                   >{\raggedright\arraybackslash}p{0.29\linewidth}
                   >{\raggedright\arraybackslash}p{0.55\linewidth}@{}}
  \toprule
  \normalfont\sffamily\bfseries\color{kzink}Exp. & \sffamily\bfseries Change & \sffamily\bfseries What we learned \\
  \midrule
  \endfirsthead
  \toprule
  \normalfont\sffamily\bfseries\color{kzink}Exp. & \sffamily\bfseries Change & \sffamily\bfseries What we learned \\
  \midrule
  \endhead
  \bottomrule
  \endfoot

  \multicolumn{3}{@{}l}{\sffamily\itshape\color{kzgrey} Stage 1 with WavLM features (prototype)} \\*
  01 & Conformer + powerset + global window linking, energy labels
     & Activity F1 0.83--0.88; only 54--64\% of conversations without swaps. \\
  02 & Same model, reference activity instead of predicted (no training)
     & No swaps at all: the bottleneck is activity, broken by crosstalk in Open Yap's energy-based labels. \\
  03 & Word-gated labels + profile heads (TS-VAD style)
     & Test clean F1 0.943, replies 0.902, 97.5\% without swaps; on real dialogues on par with pyannote. \\
  16 & Stage-1 heads fine-tuned on a noisy, reverberant chain
     & No change (noise 0.829 vs.\ 0.827): the frozen encoder's features, not the heads, lose replies in noise. \\
  17 & Activity threshold chosen on val
     & Threshold 0.4: replies +0.006 clean / +0.019 rec at $\leq$\,0.4\db{} more leak. \\

  \multicolumn{3}{@{}l}{\sffamily\itshape\color{kzgrey} Stage 2: separator} \\*
  04 & TF-GridNet S (2\,M) from scratch, conditioned on profiles and hints
     & Learns to route solo speech, does not separate overlaps (SI-SDRi $\leq$ +0.4\db{}). \\
  05 & + overlap loss term, batch 16
     & Still no separation within 40\,k windows: natural overlaps are too short and rare. \\
  06 & Word-gated labels for references
     & Not run: already in use; $-16$\db{} crosstalk in references is real speech in the headset. \\
  07 & KezekTSE-L M (13\,M), clean Libri-Light overlaps
     & +6.9\db{} at 15\,k steps, Open Yap zero-shot +2.7\db{}; stopped by an over-strict solo gate. \\
  08 & Continued, relaxed solo-suppression gate
     & +9.7\db{} on synthetic overlaps at 50\,k; Open Yap zero-shot +5.3 clean / +3.5 rec. \\
  09 & Open Yap: synthetic overlaps from Open Yap solos + natural windows + Libri replay
     & Pipeline +9.5 clean / +7.0 rec on overlaps, solo 23\db{}, leak $-21$\db{}, replies 0.90. \\
  10 & L (19.6\,M) from scratch on the full mixture
     & Too slow: +1.86\db{} rec at 12.5\,k steps. Capacity cannot be tested from scratch in budget. \\
  11 & L widened from the M weights
     & Pipeline +9.86 clean / +7.65 rec (M: +9.49 / +7.01), leak $-22$ / $-21$\db{}. Kept. \\
  12 & Short-reply loss weight $\times$3
     & Replies 0.877 / 0.825 vs.\ 0.89 / 0.838, overlaps worse: not a loss-weight problem. \\
  13 & Synthetic quiet replies
     & Not run: with reference activity the separator already keeps 91--92\% of replies, so stage 1 loses most of them. Later qualified by 49--51: replies more than 10\db{} quieter are lost by the separator too. \\
  14 & Prototype pipeline on Open Yap test
     & Clean: overlaps +10.18\db{}, solo 21.7\db{}, leak $-20.1$\db{}, replies 0.902; rec replies 0.83--0.84 in noise and reverberation. \\
  18 & Harder recording chain in synthetic data
     & No gain on rec (+7.70 vs.\ +7.73\db{}). \\

  \multicolumn{3}{@{}l}{\sffamily\itshape\color{kzgrey} Own speaker fingerprint (replacing ECAPA)} \\*
  25 & 10\,M raw-audio embedding, 7.3\,k Libri-Light voices + Open Yap, sibling classes
     & Better than ECAPA on reading and hard pairs, 4.6$\times$ faster, worse on conversational Open Yap (15.3\% vs.\ 11.7\% EER). \\
  27 & + full recording chain, 25\% conversational anchors
     & With effects applied one at a time, on par with or better than ECAPA under each except the telephone band (11.5\% vs.\ 9.1\%). \\
  28 & Telephone band on 35\% of crops
     & Better on its own band shape, worse on another shape (14.7\% vs.\ 10.8\%): needs random shapes. \\
  29 & Telephone band of random shape
     & Finds the real cause: phase. Same band without phase shift 8.5\% vs.\ 9.2\%; with IIR phase 14.1\% vs.\ 9.1\%; pure all-pass 9.6\% vs.\ 7.5\%. \\
  30 & Random all-pass cascades on 60\% of crops
     & Phase sensitivity mostly removed (mixed telephone 11.2\% vs.\ 10.8\%); Libri 5.2\% vs.\ 6.8\%; Open Yap 13.6\% vs.\ 11.7\%. \\
  31 & Phase-blind input (learned complex filterbank, modulus)
     & Telephone 10.3\% but coarser on clean speech (9.5\% vs.\ 7.5\%). \\
  33 & + 1\,558 VoxPopuli voices (116\,h)
     & Open Yap unchanged (13.9\%): the gap is not the number of voices; Libri 5.0\%; best on very similar voices. Kept. \\

  \multicolumn{3}{@{}l}{\sffamily\itshape\color{kzgrey} Version 1: own components (no WavLM, no ECAPA)} \\*
  32 & PairNet (11.7\,M): two voices for the whole recording, no window linking
     & Better activity than the prototype in every condition and $\sim$90$\times$ faster, but misses replies (0.78 vs.\ 0.89 rec). \\
  34 & Replies weighted $\times$4, twice as many synthetic replies
     & Replies 0.78 $\rightarrow$ 0.86 rec (prototype 0.89) at almost no F1 cost. \\
  35 & Separator fine-tuned on our fingerprint profiles
     & Version 1 reaches $\geq$ +10\db{} on clean val for the first time; rec on par with the prototype (+7.7\db{}). \\
  36 & Fingerprint cosines in the PairNet decoder
     & Higher F1, replies $-0.01$; resolves very similar voices that every earlier variant swapped. \\
  37 & 12\,k more steps, replies weighted $\times$6
     & Stage-1 replies stay $\leq$ 0.82 rec; pipeline replies rise only at a leak cost. \\
  38 & Version 1 vs.\ prototype on Open Yap test
     & Clean overlaps +10.49 vs.\ +10.08\db{}, 2$\times$ faster (231 vs.\ 118\,h/GPU-h); rec worse (leak $-15.6$ vs.\ $-17.0$\db{}, replies 0.81 vs.\ 0.87). \\
  39 & Separator trained on real PairNet hints, leak weight $\times$3.3
     & Leak +0.35\db{} better at the cost of overlaps ($-0.6$\db{}) and replies ($-0.04$). \\
  40 & Separator L $\rightarrow$ XL (26\,M)
     & +0.15\db{} on windows, nothing end to end. Capacity is not the lever. \\
  41 & Diagnosis: prototype stage 1 under our separator
     & Our separator is not worse; stage 1 loses replies ($-0.045$ clean), and missed replies are absent, not under threshold. \\
  42 & Recover replies with a hint-free separator pass
     & +0.007 replies, more leak: missed replies cannot be recovered this way. Later qualified by 49--51: quiet replies are also lost by the separator. \\
  43 & PairNet + log-mel branch, crops around real replies
     & Replies +1.3--3 points (over speech 0.836 $\rightarrow$ 0.853, prototype 0.876; in noise better than the prototype). Kept. \\
  44 & Continued, 70\% of crops around real replies
     & No gain: stage 1 saturates at $\sim$0.87 on rec. \\
  46 & Self-supervised pre-training of the stage-1 encoder ($\sim$900\,h)
     & Learns to detect a second voice (0.90) but does not help stage 1 (replies 0.75--0.81 vs.\ 0.82--0.85). \\

  \multicolumn{3}{@{}l}{\sffamily\itshape\color{kzgrey} External comparison and the remaining gap} \\*
  48 & Version 1 vs.\ pyannote $\rightarrow$ DialogueSidon, Open Yap test
     & Replies 0.830 vs.\ 0.687, leak $-13.9$ vs.\ $-10.1$\db{}, no-swap 0.934 vs.\ 0.793, coverage 1.0 vs.\ 0.735, 231 vs.\ 51\,h/GPU-h. \\
  49 & Separator without activity hints
     & Worse on every count (replies 0.662 vs.\ 0.708, leak $-11.2$ vs.\ $-19.9$\db{}, swaps in 4 of 20). \\
  50 & Ideal-mask ceiling for quiet replies
     & Ideal mask keeps 0.743 of replies $>$10\db{} quieter; version 1 0.171. \\
  51 & Synthetic replies 5--25\db{} below the main voice
     & Quiet replies 0.371 $\rightarrow$ 0.397 on all 113 val conversations (ideal mask 0.788 there), +0.026 only: the level of replies in training data is not the cause. \\
  52 & Version 1 vs.\ pyannote $\rightarrow$ DialogueSidon, CallHome English
     & Replies over speech 0.796 vs.\ 0.711, leak $-11.8$ vs.\ $-9.7$\db{}, no-swap 0.859 vs.\ 0.797, 232 vs.\ 46\,h/GPU-h. \\
  57 & Stage 1 as a diarizer vs.\ pyannote community-1 (DER)
     & DER 15.3 vs.\ 25.1 (Open Yap clean), 19.3 vs.\ 25.8 (as recorded), 30.1 vs.\ 18.4 (CallHome); confusion 0.5--2.0\% vs.\ 3.0--4.4\%; errors are mostly false alarm from the low threshold. \\
\end{longtable}
}

\end{document}